# Resonant surface plasmons of a metal nanosphere can be considered in the way of propagating surface plasmons

Yurui Fang[1, *] and Xiaorui Tian[2]

[1]Bionanophotonics, Department of Applied Physics, Chalmers University of Technology, SE-412 96, Göteborg, Sweden

[2]Division of Physics and Applied Physics, School of Physical and Mathematical Sciences, Nanyang Technological University, 21 Nanyang Link, 637371, Singapore

[*]Corresponding author: yurui.fang@chalmers.se.

Assuming that the resonant surface plasmons on a spherical nanoparticle is formed by standing waves of two counter-propagating surface plasmon waves along the surface, by using Mie theory simulation, we find that the dispersions of surface plasmon resonant modes supported by silver nanospheres match that of the surface plasmons on a semi-infinite medium-silver interface very well. This suggests that the resonant surface plasmons of a metal nanosphere can be treated as a propagating surface plasmon wave.

**Key words**: Surface plasmons, Surface waves, Mie theory, Dispersion relation

## Introduction

Surface plasmon polaritons (SPPs) are electromagnetic (EM) excitations propagating at the interface between a conductor and a dielectric material, evanescently confined in the perpendicular direction. These EM surface waves arise via the coupling of the EM fields to oscillations of the conductor electron plasma. SPPs are usually divided into two different kinds, which are localized surface plasmons and propagating surface plasmons, determined by if the wavevector has a real part. Localized multiple surface plasmon resonances in one dimension metallic nanostructures like nanowires, nano rods and nano rices, can be considered as Fabry-Pérot resonances of the propagating SPPs, which has been studied intensively[1-6]. Several studies were proposed on the surface plasmon dispersion relation for spherical metal nanoparticles or curved metal-dielectric interface. In 1978, Ogale et al. solved surface plasmon resonant frequencies versus radius of spherical metal nanoparticles from 0.5nm to 6nm[7]. However, the results only showed the quantum effect in very small particles (clusters) which was confirmed by recent studies[8]. Quasi-analytical method was proposed to study the wave propagating along a generally curved smooth interface of metal and dielectric medium[9]. Liaw et al. recently studied the surface plasmon waves propagating along a curved metal-dielectric interface and big metallic nanoparticles (size scale bigger than 400nm), and got dispersion relations of the surface plasmon waves creeping along a curved interface[10, 11]. However, as is usually considered, big nanoparticles originally support propagating surface plasmon waves. Later, Guasoni's excellent study extended the propagating surface plasmon waves to a 200nm diameter nanoparticle which is usually considered to only support localized surface plasmons[12]. Nordlander et al. also got the plasmon dispersion relation for a planar thin metal film from the plasmon resonances of a metallic nanoshell with a limit of infinite radius[13]. In this paper, we extended this further. With Mie theory simulation of silver spherical nanoparticles, the dispersion relation of the SPPs is obtained and analyzed. It is found that the dispersion curves of localized SPPs on a spherical nanoparticle excellently match with that of propagating SPPs on the plane interface of metal and dielectric medium, where the radius of the nanoparticle is finite, from several hundreds of nanometers to 10nm.

## Assumption

First we define that when a spherical metal nanoparticle with radius r embedded in homogeneous medium is resonant at certain wavelengths, the orders (dipole or multiples) of different resonant modes are expressed with m = 1 (first order, dipole), 2 (second order,), 3 (third order), ..., as shown in Fig. 1a and 1b. Then we assume that for each order, when the surface plasmon resonance happens, it is a standing wave formed by two SPP waves propagating clockwise and counterclockwise along the surface. So the wavelength of the wave will be

$$\lambda_m = 2\pi r/m \quad (1).$$

When m = 1 (dipole resonance), the wavelength of the surface plasmons equals to the perimeter of the sphere, i.e. $2\pi r$. We also assume that the resonant surface plasmons are equivalent to that with the same wavelength propagating on the semi-infinity silver-air interface (Fig. 1c). Thus the wave vector of the SPPs on sphere can be defined as

$$k_m = \frac{2\pi}{\lambda_m} = \frac{m}{r} \quad (2).$$

From the resonant peak and wave vector, the dispersion relation of the spherical particles can be deduced.

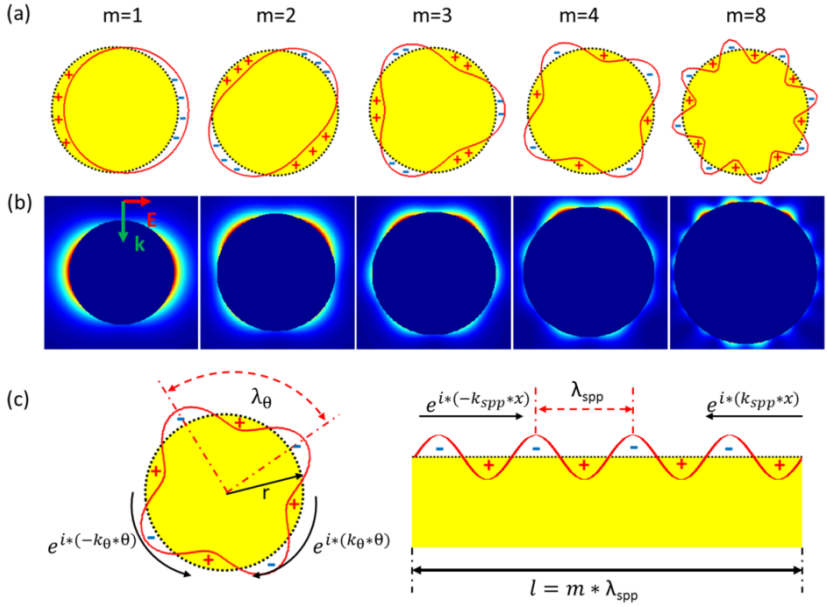

Figure 1. (a) Illustration of charge distributions of different orders of surface plasmon resonant modes. (b) Corresponding electric field patterns of a silver nanosphere (Drude model, $\varepsilon = 3.7 - \frac{\omega_p^2}{\omega^2 + i*\omega*\frac{1}{\tau}}$, $\omega_p$ = 9.149 eV, τ = 3.1e-14 s) in air calculated by Mie theory (from left to right, R = 20 nm, λ = 335 nm; R = 30 nm, λ = 315 nm; R = 55 nm, λ = 310 nm; R = 100 nm, λ = 315nm; R = 300 nm, λ = 335 nm). (c) Illustration of the assumption of a surface plasmon resonant mode of spherical nanoparticle and surface plasmons on the interface of plane metal and dielectric medium. The surface charges of an m = 4 order surface plasmon resonance can be viewed as a standing wave of a SPPs wavelength $\lambda_\theta$, which is formed as two waves (clockwise and counterclockwise). It is equivalent to the surface charge distribution of a surface plasmon wave on a plane interface of the metal and dielectric medium.

## Results and discussion

Based on the assumption above, silver (Johnson & Christy experiment[14]) spherical nanoparticles in air are calculated with Mie theory (and verified with full wave calculation by using FEM method). Fig. 2a shows the extinction spectra of nanospheres in air with diameters from 10 nm to 380 nm. The excitation light is plane wave and the polarization is fixed in one direction. For metal particles bigger than 10 nm, the quantum effect is negligible and one also the dielectric function is still usable without further modification (simulations with electon mean free path modified dielectric function show very little difference). Wavelengths and thus frequencies of surface plasmon resonant peaks of different orders (m = 1, 2, 3, …) can be obtained from the spectra; wave vectors are deduced from formula (2). Then the SPPs dispersion relations of the nanospheres are plotted in Fig. 2b, together with the dispersion relation of the SPPs on the interface of semi-infinite silver-air interface (TM waves). It can be seen that the dispersion relations of all of the orders of SPP resonant modes on the nanosphere, match very well with that of the SPPs on a planar surface. This suggests that our assumption is correct, i.e. the localized SPPs on a spherical nanoparticle actually are kinds of propagating SPP waves. However, in the range of large wave vectors, there are still some deviations between the dispersion relation curves of the nanosphere and that of the planar film, with different deviations for different orders of modes. All modes of the nanosphere have lower energy than that of the SPPs in the planar film, which is corresponding to the electrostatic approximation of spherical SPPs.

To further check the relationship between the localized SPPs on spherical nanoparticles and propagating SPPs on planar surface, Mie theoretical simulation are performed under Drude Model. The dielectric function is set as

$$\varepsilon = 3.7 - \frac{\omega_p^2}{\omega^2 + i*\omega*\frac{1}{\tau}} \quad (3),$$

where $\omega_p$ = 9.149 eV, $\tau$ = 3.1×10$^{14}$ s ($\varepsilon_\infty$ = 3.7 is used as the permittivity to make the value more close to the real metal). With the same method stated above, the dispersion relations are plotted in Fig. 3. Figure 3a shows the dispersion relations of nanospheres with the different radius, as well as the dispersion relations of SPPs on semi-infinite silver-air interface (dark cyan) and the light line in air. Different numbers refer to different resonant orders (m = 1, 2, 3, …) of plasmonic modes. It can be seen that the dispersion relation lines of every order mode match very well with the one of SPPs on silver-air interface. In the range of large wave vectors, the dipole modes have the lowest energy. Higher order modes have gradually increasing energies, which is consistent to the resonant situation of spherical nanoparticles under the electrostatic approximation ($\omega_{sp,l} = \omega_p/\sqrt{l/(2l+1)}$). As the wave vector becomes smaller, the dispersion relation curves of all modes meet together with the one of SPPs on silver-air interface. The deviations also show that the surface plasmon waves propagating on the spherical nanoparticles have a smaller phase velocity, which was confirmed in Liaw's work[11]. When the radius of curvature becomes bigger, the phase velocity is more close to the one of planar surface. Figure 3b shows the situation of big nanoparticles, in which an excellent matching is obtained.

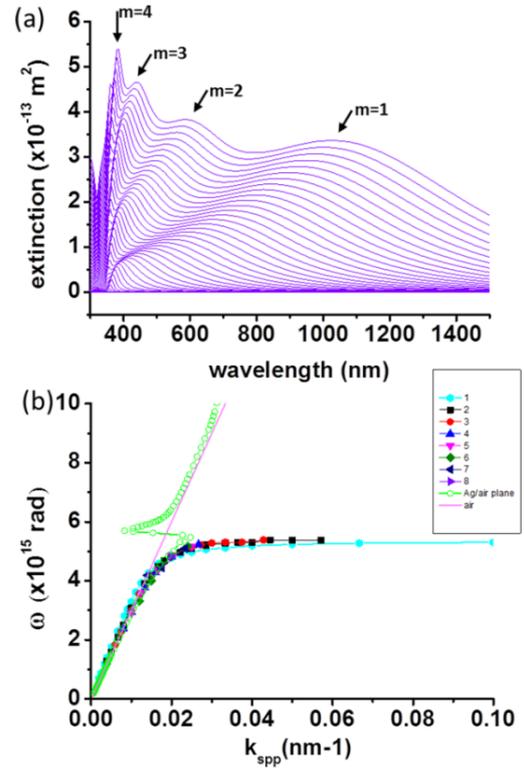

Figure 2. (a) Mie theory calculated extinction spectra of silver nanosphere in air with radius from 5 nm to 190 nm. (b) Dispersion relation of the SPPs of the spherical nanoparticles (same with (a)) and silver planar surface.

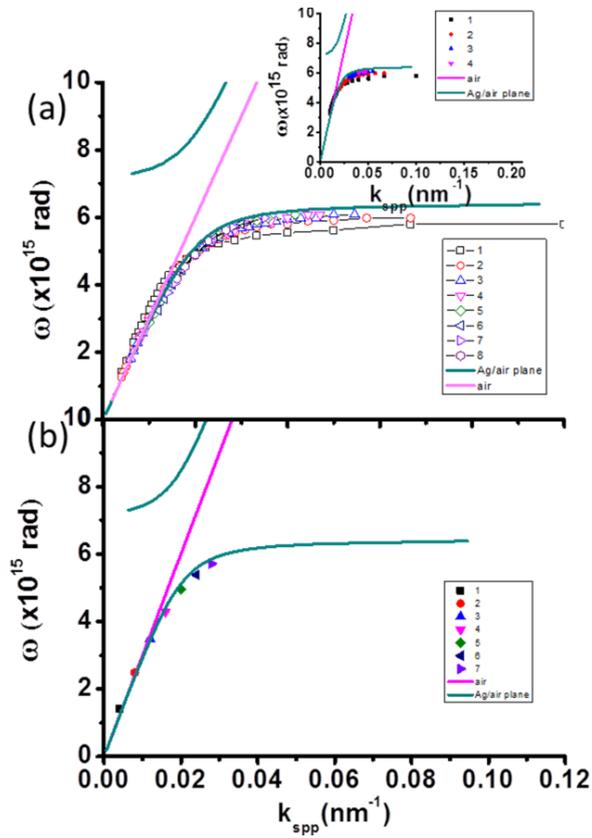

**Figure 3.** Dispersion relations of SPPs on silver spherical nanoparticles in air and that of SPPs in the interface of air and silver planar surface, obtained from Drude mode. The parameters are the same as in Fig. 1. (a) The nanoparticle radius is from 5 nm to 500 nm. The inset shows the dispersion relations of nanoparticles with radius from 5 nm to 100 nm. (b) Only the dispersion relation of the nanoparticle with R = 250 nm is plotted.

As analyzed above, the localized surface plasmon resonance can be considered as two anti-propagating surface plasmon waves which form a standing wave. It not only applies to the high order modes of big nanoparticles, but also applies to small particles and dipole modes. However, the conclusion doesn't apply to infinite metal cylinder. The reason may be that the cylinder and sphere are different classes of structure in plasmonics as shown in transform optics[15, 16]. And the results from both Schmidt and Chang also show that the SPPs on metal cylinder have mixed wave vector in both angular in cross section and along the cylinder directions[17, 18]. It should be noted that even in the range of small wave vectors, the dispersion relation curve of dipole modes (but not for the higher modes) still deviates a little bit from the one of SPPs on silver-air interface, though the deviation is so small and not obvious. It may be because for big particles, the retardation effect cannot be ignored and thus the resonant frequency blue shifts.

### Conclusion

In conclusion, under the assumption of surface plasmon waves on a spherical nanoparticle analogous to the propagating ones on a planar surface, we plot the dispersion relations of surface plasmon resonances of spherical nanoparticles. The results show that the dispersion relations of SPPs on a spherical particle excellently match that of the surface plasmons on a planar surface. This means that the localized SPPs can be considered in a way that two anti-propagating surface plasmon waves propagate on the surface particles and form a standing wave. The conclusion is still applicative even for small nanoparticles with radius smaller than 200 nm until to 5 nm, and for dipole resonance as well. This gives us another perspective on localized surface plasmon polaritons.

### Acknowledgements:

Yurui would like to thank useful discussion with Mikael Käll and Peter Johansson.